\documentclass[12pt]{article}
\usepackage{amsmath}
\usepackage{amssymb}
\usepackage{multirow}
\usepackage{graphicx}
\usepackage{setspace,caption}
\usepackage{lipsum}
\usepackage[section]{placeins}
\usepackage[T1]{fontenc}
\usepackage{lscape}

\usepackage{latexsym}

\begin{document}


\title{
 Stock Trading Using PE ratio: A Dynamic Bayesian Network Modeling on Behavioral Finance and Fundamental Investment
}

\author{Haizhen Wang, Ratthachat Chatpatanasiri, Pairote Sattayatham \\
	School of Mathematics, Suranaree University of Technology,  THAILAND\\
    \texttt{wanghaizhena@163.com, ratthachat.c@gmail.com, pairote@sut.ac.th}
}

\maketitle

\begin{abstract}
On a daily investment decision in a security market, the \textit{price earnings (PE) ratio} is one of the most widely applied methods being used as a firm valuation tool by investment experts. Unfortunately,  recent academic developments in financial econometrics and machine learning rarely look at this tool. In practice, fundamental PE ratios are often estimated only by subjective expert opinions.  The purpose of this research is to \textit{formalize a process of fundamental PE estimation} by employing advanced dynamic Bayesian network (DBN) methodology. The estimated PE ratio from our model can be used either as a information support for an expert to make investment decisions, or as an automatic trading system illustrated in experiments. Forward-backward inference and EM parameter estimation algorithms are derived with respect to the proposed DBN structure. Unlike existing works in literatures, the economic interpretation of our DBN model is well-justified by behavioral finance evidences of volatility. A simple but practical trading strategy is invented based on the result of Bayesian inference. Extensive experiments show that our trading strategy equipped with the inferenced PE ratios consistently outperforms standard investment benchmarks.
\end{abstract}


\section{Introduction}\label{sec1}

With the rapid advancement of machine learning technology, recent works make attempts to incorporate these machine learning techniques to construct trading systems that support decisions of investors in security markets \cite{YehEtal2010,Luetal2009,WenEtal2010,Hassan2009,KaoEtal2013,KazemEtal2013} (see also references therein). These recent works have the following common philosophical theme. \\

\noindent \textbf{Common Philosophy of Applying Machine Learning to Financial Data: } \\
There exist \textit{hidden patterns} in  financial time series. Complicated techniques (and their combinations) such as support vector machine, multiple kernel learning, independent component analysis, hidden Markov models, fuzzy modeling and so on, can help investors discover \textit{hidden patterns represented by complicated mathematical formulas}. The retrieved formulas can be used as forecasting rules to predict stock's directional movements, which in turn can be incorporated into investor's trading strategy (buy low and sell high, as predicted by the rules) to make an excess return in the market.\\

Based on the same philosophy mentioned above, existing works, however, have common limitations. Firstly, the discovered patterns are so complicated (highly non-linear) and lacked of financial interpretation; note that, in general, \textit{higher degree of pattern complexities are more risky to overfit the training data \cite{bk:Bishop2006}.} Secondly, each financial time-series has to be trained separately, resulting in one set of distinct patterns for each different security. In other words,\textit{ there is no common pattern in the data of interested securities}. Thirdly,  because of pattern complexities, practical trading implementations are not easy for some investors. In fact, sophisticated trading program has to be constructed by users themselves. Lastly, there is no direct way to incorporate existing expert information (such as professional security analysts' recommendations) into the learning system. Fairly speaking, although having the mentioned limitations, the core philosophy of existing research matches the philosophy of one certain investor group called \textit{technical analysts} \cite{bk:Murphy1999,bk:Shannon2008}. Technical analysts believe in price patterns and do not pay much attention to economic interpretation of the patterns. Therefore, this line of existing research may benefit this group of investors.

On another side of investment practitioners, there is a group named \textit{fundamentalists} whose trading strategies have clear financial interpretations and are based on well-defined financial information \cite{bk:Mark2011,bk:Damodaran2012,bk:LynchRothchild2000}. \textit{Price-earning ratio} (simply called \textit{PE ratio}, to be defined below shortly) is one of the most widely applied valuation toolkits for fundamentalists to make their investment decisions \cite{bk:Damodaran2012,bk:HenryEtal2010}. Also, investment recommendations by security analysts are often based on PE ratios \cite{bk:CarvellEtal1989}. Nevertheless, it is unfortunate that recent academic advancements in financial econometrics and machine learning rarely look at this tool so that the current practical application of PE ratio has to depend solely on expert knowledge. To our knowledge, there is currently no formal framework capable of integrating expert knowledge with historical financial time-series data to make a systematic inference of PE ratio from available information.

In this research, we focus on applying a Bayesian statistical analysis to \textit{formalize the process of stock valuation using the PE ratio}. We apply the powerful framework of \textit{dynamic Bayesian network} \cite{bk:Bishop2006,bk:Murphy2012} to model the valuation process. In contrast to existing machine learning frameworks mentioned above on price pattern discovery where the discovered patterns have no meaning in finance, the interpretation of our model is well justified according to \textit{behavioral finance} \cite{bk:Szyszka2013} as explained in the Section  \ref{sec2}. The main contributions of our work are threefold. Firstly, to our knowledge, we are the first to propose applying the machine learning framework to formalize the PE ratio valuation process which somehow rarely gets attention from academic researchers. Secondly, unlike existing works where there are different discovered patterns for different securities, our proposed trading strategy resulted by the Bayesian framework is unified, i.e. we propose a single trading strategy which can be applied to every security as explained in Subsection \ref{subsec1.1}. The proposed strategy  is simple and has a clear financial interpretation so that it can be easily applied by every practitioner (no requirement on writing a sophisticated system trading program). Moreover, expert opinions can be naturally integrated to our Bayesian learning framework. Thirdly, as our proposed dynamic Bayesian network having non-standard structure compared to literatures \cite{bk:Bishop2006,bk:Murphy2012}, we have successfully derived the new inference formulas by applying the forward-backward methodology, and the new parameter estimation algorithm according to the concept of Expectation-Maximization \cite{bk:Bishop2006,bk:Murphy2012}.

Note that in this paper, we focus on investment in individual firm-level securities which are usually preferred by individual investors; in contrast to investment institutions whose investment strategy is usually on portfolio level based on \textit{Modern Portfolio Theory} \cite{Barber2011}.

\subsection{Background of fundamental investment based on PE ratio}\label{subsec1.1}

The core idea of the PE ratio valuation method is simply that\textit{ the value of the firm (and hence the value of its stock) is directly proportional to the annual net income (also called earning) of the company}, i.e. for each firm $i$,
\begin{equation}\label{Lu}
P_{i}^{*}=PE_{i}^{*}\times E_{i}
\end{equation}
where $P_{i}^{*}$    denotes the value of firm $i$ ,
 $E_{i}$ denotes the firm's current annual earnings and
$PE_{i}^{*}$ is the firm's appropriate PE ratio, usually assumed to be a constant (at least for some period of time). Here, the annual earning is defined by the summation of the latest four quarterly earnings. The earnings information of each firm listed in the stock market is normally available to all investors, i.e., it is observable.

The PE ratio, intuitively, can be thought of as a\textit{ premium} of an individual firm, i.e., given the same earning for two firms, the firm with higher PE ratio is considered to be of higher value. Conceptually, the appropriate PE ratio of each firm is usually determined by experts using\textit{ business and financial accounting factors} such as debt burden, cash flow, growth rate, business risk, etc.  There exists an alternative approach to estimate the PE ratio called the \textit{relative approach}  \cite{bk:Damodaran2012} which still requires experts to select a group of similar firms altogether, and the ratio is heuristically calculated from this group. To summarize, \textit{the current best practice for PE ratio estimation is to be heuristically calculated by experts or experienced investors}.

Once we get the PE ratio, we can simply calculate firm value, often called \textit{intrinsic value}, by Eq.(\ref{Lu}). A simple trading strategy is to compare the firm value with a market price of the firm.

\underline{\textbf{Strategy A}}: if the firm value is higher than its market price by some threshold, it is considered to be at low price, so that we can buy the firm's stock. We expect to sell it later when its market price is higher than the firm's intrinsic value by some threshold.

 It is important to note that the philosophy of this trading strategy is that \textit{market price is not always equal to the value of the firm}. We can observe that the price of firm's stock changes almost every working day in a stock market. In contrast, by
 Eq. (\ref{Lu}), the firm's value will not change in a short time period provided that there is no new announcement on annual earnings in that period. There has been long controversy about this \textit{“price vs. value”} issue \cite{bk:Mark2011}, but it is beyond the scope of this paper. In any cases, it is a fact that there exists a large group of individual investors namely \textit{fundamentalists} employing the PE ratio as their main tool. Instead of solely relying on expert opinions, the goal of this paper is to support that group of investors to \emph{systematically} determine the appropriate PE ratio, by the method of Bayesian statistical analysis which is able to formally combine information from historical data with expert beliefs.

 Finally, we emphasize that there is another quantity called an \emph{observed PE} calculated from firm's current market price divided by its earnings (note again difference between \emph{value} $P^*_{i}$ and \emph{price} $P_{i}$), that is,

 \begin{equation}
 observed\ PE_{i}=P_{i}/E_{i},
\end{equation}

 where $P_{i}$ is the current market price of a firm $i$. In Bayesian analysis of the PE ratio, it is important to distinguish between the observed $PE_{i}$ (changing everyday due to changes of $P_{i}$) and $PE_{i}^{*}$. Here we will call $PE_{i}^{*}$ as the \emph{fundamental PE ratio}. The reason behind this name is the following: only a group of fundamentalists believe that the quantity $P_{i}^{*}$ , or firm value, is able to calculated by Eq. (\ref{Lu}). Therefore, they usually call $P_{i}^{*}$ as the \emph{fundamental price} or \emph{fundamental value}, and so $PE_{i}^{*}$ as \emph{fundamental PE}. To them, there exist various kinds of investors in the market: \emph{some are rational and some are irrational}. The current market price $P_{i}$ and hence, too, the \emph{observed} $PE_{i}$ can fluctuate from the fundamental price $P_{i}^{*}$ by actions of those irrational investors. We shall formally model this argument in Section \ref{sec2}.

\section{Statistical model of stock price dynamics}\label{sec2}

\subsection{Motivation of statistical modelling: behavioral volatility}\label{subsec2.1}

In Subsection \ref{subsec1.1} we mentioned about fundamentalists' belief that market price of a security may not equal to its fundamental value.  Why does stock price deviates from its fundamental price? Works on behavioral finance \cite{bk:Szyszka2013} found many evidences for this question. For example, researchers argue that there are \textit{noise traders} in the market who tend to make irrational actions so the price can move away from its value \cite{Black1986,DeLongEtal1990,bk:Hommes2013}. One of the works found that some investors cannot process new information correctly and so \textit{overreact} to new information \cite{DeBondtThaler1986}. What is worse, information which investors overreact is, many times, unconfirmed \cite{BloomfieldEtal2000} or unreliable \cite{PoundZeckhauser1990,TumarkinWhitelaw2001} or even unimportant \cite{Rashes2001,CooperEtal2001}. Also, investors who consult experts may not get much helpful advice since security analysts tend to be overoptimistic \cite{DechowEtal2000} and having conflict of interest \cite{CowenEtal2006}. Finally, it is well known that even rational investors  in the market cannot immediately eliminate this irrational pricing due to \textit{limit of arbitrage} \cite{ShleiferVishny1997}. All the effects mentioned here are able to temporarily move away a stock price from its value for a period of time. This is what we call \emph{behavioral volatility}. The effects continue until either they are cancelled out, or rational investors finally eliminate this mispricing. This reversion phenomena is called \textit{mean reversion} in literatures.

\subsection{Dynamic Bayesian Network of stock price movement}\label{subsec2.2}

Our model simplifies and formalizes the observations described in Subsection \ref{subsec2.1}. We divide the temporary effects which cause mispricing into two categories: (1) \textit{short-term effects}: mispricing effects which last about a few days, e.g. effects causing by noise trading or overreaction to unreliable information and (2) \textit{medium-term effects} : mispricing effects which last several weeks or months, e.g. effects caused by reaction to unconfirmed information which may take time to confirm, or overoptimistic prediction of analysts which may take time to prove. Mathematically, the relation between market price and its fundamental value can be described as the following equation. To simplify the equation, since we consider only one firm at a time, we now replace the firm-index subscript $i$ with a time-index subscript $t$ to emphasize the dynamic relationship between price and its fundamental value.
\begin{equation}\label{a}
  P_{t}=P_{t}^{*}(1+z_{t})(1+\varepsilon_{t})
\end{equation}
where

(a) $z_{t}$ is a random variable modeling the medium-term noisy effects. To make its effects persist for a period of time, we model ${z_{t}}$ as a Markov chain.

(b) $\varepsilon_{t}$ is a random variable for the short-term noisy effects which is modeled by a Gaussian random noise, $\varepsilon_{t}\sim N(0,\sigma^{2})$.

\begin{figure}[!htb]
	\centering
	
	\includegraphics[width=\linewidth]{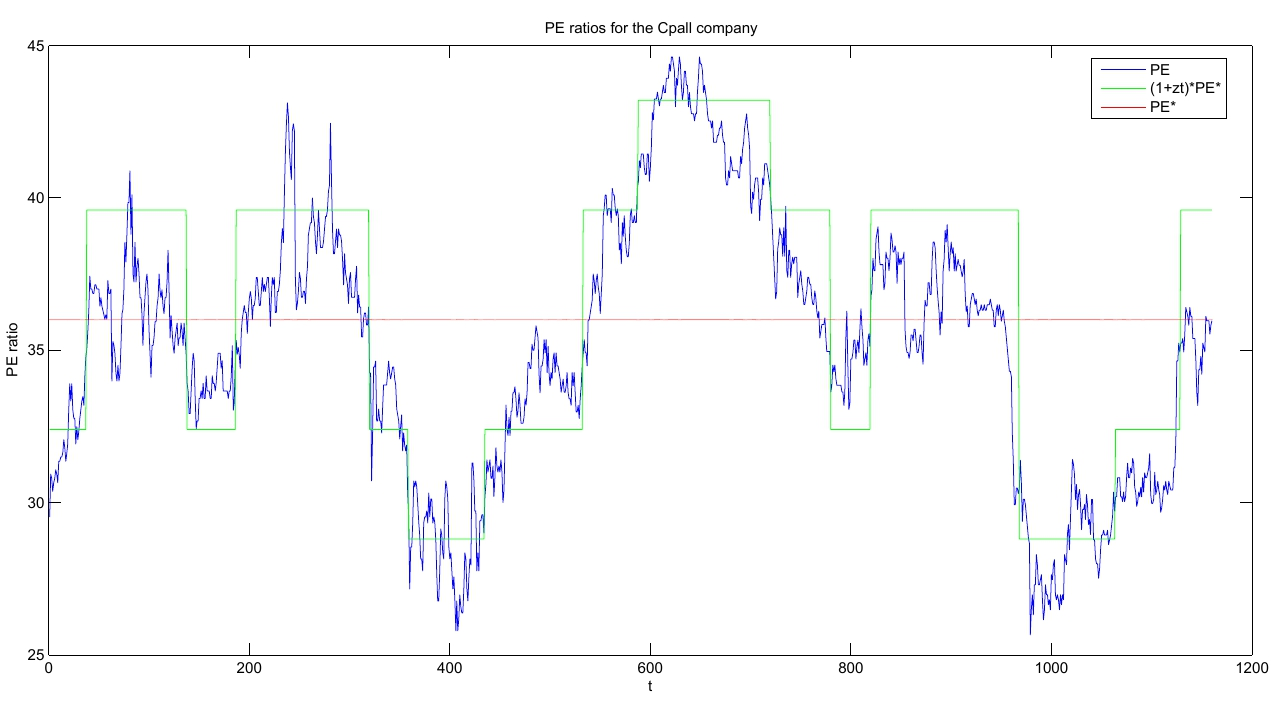}
	
	\caption  {Illustration of our main idea described by Eq. (\ref{c}). The plot is the observed PE vs. date. Red dashed line shows $PE^{*}$ of the firm, while Green line illustrates the effect of medium-term noisy effect $z_{t}$ to $PE^{*}$. Blue line illustrates the observed PE which is affected by both the medium-term and short-term mispricing effects.}
	
	\label{fig:Fig1}
	
\end{figure}

Assuming $PE^{*}$ as a constant for the period which we observed, and following Eq. (\ref{Lu}) of Section \ref{sec1}, we have
\begin{equation}\label{b}
   P_{t}=PE^{*}E_{t}(1+z_{t})(1+\varepsilon_{t})
\end{equation}
and, therefore, we get the relationship between the fundamental PE and the observed PE:
\begin{equation}\label{c}
 P_{t}/E_{t}=PE^{*}(1+z_{t})(1+\varepsilon_{t})
\end{equation}
Note that our model is suitable only for a firm with positive earnings $E_{t}>0$. Fortunately, most firms satisfy this criterion. Eq. (\ref{c}) is central to our idea and can be visualized as shown in Figure \ref{fig:Fig1}.

We can mathematically simplify Eq. (\ref{c}) further
\begin{equation}\label{d}
  \ln(P_{t}/E_{t})=\ln(PE^{*}(1+z_{t}))+\ln(1+\varepsilon_{t})
\end{equation}

\begin{figure}[!htb]
  \centering
  \includegraphics[width=0.6\linewidth]{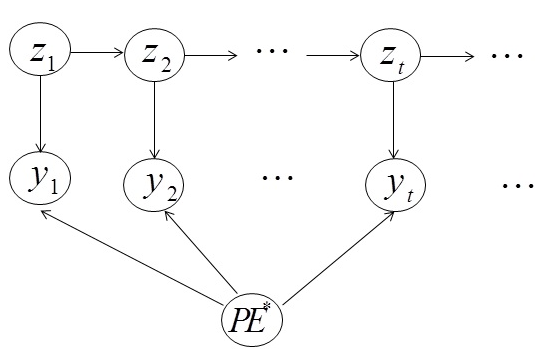}
  \caption{The proposed model represented by dynamic Bayesian network (DBN). $y_{t}$ is an observable quantity, while $PE^{*}$ and $\{z_{t}\}$ are unobservable.}
  \label{Fig2}
\end{figure}

Since $\varepsilon_{t}$ is usually small, it can be approximated by $\ln(1+\varepsilon_{t})\approx\varepsilon_{t}$, and denote $y_{t}=\ln(P_{t}/E_{t})$, we then have
\begin{equation}\label{de}
  y_{t}=\ln(PE^{*}(1+z_{t}))+\varepsilon_{t}
\end{equation}
Note that as explained in Section \ref{sec1}, $y_{t}$ is an observable quantity, while $PE^{*}$ and ${z_{t}}$ are unobservable, i.e. they are hidden state or latent variables. Note that these are two different types of latent variables, i.e. $PE^{*}$ is constant and ${z_{t}}$ is time-varying. Thus, Eq. (\ref{de}) is different from standard state-space and graphical models such as \textit{Hidden Markov Models} or \textit{Linear State Space Model}  \cite{bk:Bishop2006}. The graphical model of our proposed stock price dynamic has three layers as represented in Figure \ref{Fig2}. In our case, where the model is temporal, the graphical model framework is also called \textit{dynamic Bayesian network} (DBN). The main advantage of DBN is its ability to encode conditional independent properties, and hence simplifying probabilistic inference  \cite{bk:Murphy2012}. Another advantage of this framework is that expert knowledge can be integrated in the model naturally as shown in the next section.

To derive mathematical equations for inference and parameter estimation in the DBN framework, we shall assume that all latent random variables are discrete: $z_{t}\in{\{a_{1},...,a_{M}\}}$, $ PE^{*}\in{\{b_{1},...,b_{N}\}}$. Furthermore, we have to set up the conditional probability distribution function for each node given its parents. We define the conditional probability distribution functions of all nodes as follows:

 \noindent \textbf{The transition probability distribution function:}

Let $i,m\in {\{1,...,M\}}$, $t\in {\{2,3,...\}}$
\begin{equation}\label{ef}
  p(z_{t}=a_{i}\mid z_{t-1}=a_{m})\triangleq w_{im}.
\end{equation}
Note that $0\leq w_{im}\leq 1$,\ $\sum_{m=1}^{M} w_{im}=1 $ . The matrix $\textbf{W}=(w_{im})_{M\times M}$ is called a transition matrix, i.e. $\{z_{t}\}$ is a Markov chain.

\noindent \textbf{The emission probability distribution function:}

For all $m\in\{1,...,M\}$, $n\in {\{1,...,N\}}$,$t\in {\{1,2,...\}}$
\begin{equation}\label{f}
  p(y_{t}|z_{t}=a_{m},PE^{*}=b_{n})\triangleq\phi_{mn}(y_{t})
\end{equation}
By Eq. (\ref{d}), $\phi_{mn}(y_{t})=\mathit N(\ln(b_{n}(1+a_{m})),\sigma^{2})$. The matrix $\Phi_t=(\phi_{mn})_{M\times N}$ is called an emission matrix at period $t$.

\noindent \textbf{The inital probability distribution functions:}

For each $m\in\{1,...,M\}$,
\begin{equation}\label{g}
  u_{m}\triangleq p(z_{1}=a_{m})
\end{equation}
where $0\leq u_{m}\leq 1$ and $\sum_{m=1}^{M}u_{m}=1$.

For each $n\in\{1,...,N\}$,
\begin{equation}\label{h}
  v_{n}\triangleq p(PE^{*}=b_{n})
\end{equation}
where $0\leq v_{n}\leq 1$ and $\sum_{n=1}^{N} v_{n}=1$.

The vectors $\textbf{u}=(u_{m})_{M}$ and $\textbf{v}=(v_{n})_{N}$ are called initial vectors.

Therefore, in this Bayesian framework, the set of model parameters is $\theta=\{\textbf{W,u,v},\sigma^{2}\}$ and our parameters space is
\begin{equation}\label{aa}
  \Theta=\{\theta|0\leq u_{m}\leq1, \sum_{m=1}^{M}u_{m}=1, 0\leq v_{n}\leq1, \sum_{n=1}^{N} v_{n}=1,0\leq w_{im}\leq1,\sum_{m=1}^{M} w_{im}=1,\sigma>0\}.
\end{equation}

If we know all parameters, we can make an inference by deriving inference equations based on the \textit{forward-backward} algorithm as shown in Subsection 3.1. If the parameters are unknown, we have to estimate them first. In this paper, we derive the estimation procedures based on \textit{Maximum a Posteriori} (MAP) and \textit{Expectation-Maximization} (EM) algorithms. In the next section, we show how we derive both the inference and parameter estimation algorithms.

\section{Bayesian inference on the DBN of stock price dynamic}\label{sec3}

As explained in previous sections, our goal is to make an inference on PE* so that we can estimate the fundamental price of a stock. In Section \ref{sec4}, we will show that estimations of $\{z_{t}\}$ is also useful in investment. To infer the values of these two latent variables, similar to Hidden Markov Models (HMM) and Linear State Space Model (LSSM) \cite{bk:Bishop2006,bk:Murphy2012}, we need to derive equations in two steps. First, the inference algorithms with known parameters, second, the parameter estimation algorithms given that parameters are unknown. However, because there are two types of latent states as explained in the previous section, our graphical model shown in Figure \ref{Fig2} is more sophisticated than HMM and LSSM. In this section, we show the new equations for both inference tasks. To simplify the notation, we use notation $x_{1}^{T}$ to denote$\{x_{1},...,x_{T}\}$ .

\subsection{Inference with known parameters}\label{subsec3.1}

Suppose $\theta$ is known, together with the observed data $y_{1}^{T}$. Similar to HMM, in order to estimate the latent states of $z_{1}^{T}$ and $PE^{*}$, we need to find recurrent formulas to calculate two quantities: the \emph{filtering} probabilities   $p(z_{T},PE^{*}|y_{1}^{T},\theta)$ and the \emph{smoothing} probabilities $p(z_{t},PE^{*}|y_{1}^{T},\theta)$, $t\in{1,...,T-1}$  . To keep our formulas simple, in this Section, we will omit writing $\theta$  in the probability notations, e.g., we simply write $p(z_{T},PE^{*}|y_{1}^{T})$  for filtering.

The filtering formula, which estimates conditional joint probabilities of the \emph{most recent} medium-term effect $z_{T}=a_{m}$ and $PE^{*}=b_{n}$ given all the observed variables, is given by the following recurrent formula:
\begin{equation}\label{bb}
 \begin{array}{l}
 \indent p(z_{T}=a_{m},PE^*=b_{n}|y_{1}^{T})   \\[10pt]
   \qquad = p(z_{T}=a_{m},PE^*=b_{n}|y_{1}^{T-1},y_{T}) \\[10pt]
   \qquad \propto p(y_{T}|y_{1}^{T-1},z_{T}=a_{m},PE^*=b_{n})p(z_{T}=a_{m},PE^*=b_{n}|y_{1}^{T-1})\\[10pt]
   \qquad =\phi_{mn}(y_{T})\sum_{i=1}^{M}p(z_{T}=a_{m},z_{T-1}=a_{i},PE^*=b_{n}|y_{1}^{T-1})\\[10pt]
   \qquad =\phi_{mn}(y_{T})\sum_{i=1}^{M}  p(z_{T-1}=a_{i},PE^*=b_{n}|y_{1}^{T-1})  p(z_{T}=a_{m}|z_{T-1}=a_{i}) \\[10pt]
   \qquad =\phi_{mn}(y_{T})\sum_{i=1}^{M}  p(z_{T-1}=a_{i},PE^*=b_{n}|y_{1}^{T-1}) w_{mi}
 \end{array}
\end{equation}

In the above derivation, Bayes's rule, \emph{conditional independent} properties \cite{bk:Murphy2012} with respect to DBN shown in Figure \ref{Fig2}  and sum rule are applied consecutively to get the above result, similar to the filtering equation of HMM. The initial equation of the recurrent formula can be derived similarly: $p(z_{1}=a_{m},PE*=b_{n}|y_{1})=\phi_{mn}(y_1)u_{m}v_{n}$.

Next, the smoothing formula, which estimates conditional joint probabilities of the\emph{ any-date} $t < T$ medium-term noisy effect $z_{t}=a_{m}$  and $PE^*=b_{n}$ given all the observed variables, is given by the following so-called \emph{forward-backward} formula in Eq. (\ref{cc}) :

For all $t\in\{1,...,T-1\}$, (note that $y_{1}^{T}=y_{1}^{t}\bigcup y_{t+1}^{T})$
\begin{equation}\label{cc}
 \begin{array}{l}
 \indent p(z_{t}=a_{m},PE^{*}=b_{n}|y_{1}^{t},y_{t+1}^{T})   \\[10pt]
   \qquad \propto p(y_{t+1}^{T}|y_{1}^{t},z_{t}=a_{m},PE^{*}=b_{n})p(z_{t}=a_{m},PE^{*}=b_{n}|y_{1}^{t})\\[10pt]
  \qquad =p(y_{t+1}^{T}|z_{t}=a_{m},PE^{*}=b_{n})p(z_{t}=a_{m},PE^{*}=b_{n}|y_{1}^{t}).
 \end{array}
\end{equation}

Note that \emph{conditional independent} properties of our DBN are applied in the first term. Also note that the second term is in fact a filtering probability. Therefore, we need to concentrate only on the first term, which has the following recurrent relation:

\begin{equation}\label{dd}
 \begin{array}{l}
 \indent  p(y_{t+1}^{T}|z_{t}=a_{m},PE^{*}=b_{n})   \\[10pt]
 \qquad =\sum_{i=1}^{M} p(y_{t+1}^{T}, z_{t+1}=a_{i}|z_{t}=a_{m},PE^{*}=b_{n}) \\[10pt]
 \qquad =\sum_{i=1}^{M} p(y_{t+1}^{T}|z_{t+1}=a_{i},z_{t}=a_{m},PE^{*}=b_{n})p(z_{t+1}=a_{i}|z_{t}=a_{m},PE^{*}=b_{n})\\[10pt]
 \qquad =\sum_{i=1}^{M} p(y_{t+1}^{T}|z_{t+1}=a_{i},PE^{*}=b_{n})p(z_{t+1}=a_{i}|z_{t}=a_{m})\\[10pt]
 \qquad =\sum_{i=1}^{M} p(y_{t+1}^{T}|z_{t+1}=a_{i},PE^{*}=b_{n})w_{im} \\[10pt]
 \qquad =\sum_{i=1}^{M} p(y_{t+1},y_{t+2}^{T}|z_{t+1}=a_{i},PE^{*}=b_{n})w_{im} \\[10pt]
 \qquad =\sum_{i=1}^{M} p(y_{t+2}^{T}|y_{t+1},z_{t+1}=a_{i},PE^{*}=b_{n})p(y_{t+1}|z_{t+1}=a_{i},PE^{*}=b_{n})w_{im}\\[10pt]
 \qquad =\sum_{i=1}^{M} p(y_{t+2}^{T}|z_{t+1}=a_{i},PE^{*}=b_{n})\phi_{in}(y_{t+1})w_{im}.
 \end{array}
\end{equation}

The end condition can be solved similarly $p(y_{T}|z_{T-1}=a_{m},PE^{*}=b_{n})=\sum_{i=1}^{M} \phi_{in}(y_{T})w_{im}.$

With the derived recurrent formulas, we can get the most probable values of wanted latent variables $PE^{*}$ and each $z_{t}$ by using marginalization, e.g. \[PE^{*}=\operatorname*{arg\,max}_{b_n} p(PE^{*}=b_{n}|y_{1}^{T}),\] where $p(PE^{*}=b_{n}|y_{1}^{T})=\sum_{m=1}^{M} p(z_{t}=a_{m},PE^{*}=b_{n}|y_{1}^{T})$. To implement both filtering and smoothing in computer program, we also need to solve the formulas for the constants appeared in the above derivations. To fulfil this task, using matrix reformulation of the above recurrent equations is the most convenient and efficient way. Below, we give only the end results because the details exceed space limitation. Derivation details can be found in the full version of this paper \cite{bk:Haizhen2017}.

To get matrix formula, first denote a filtering density as $\alpha_{tmn} = p(z_{t}=a_{m},PE^*=b_{n}|y_{1}^{t})$. When $t>2$, define $\alpha'_{tmn} = \phi_{mn}(y_t)\sum_{i=1}^M w_{mi}\alpha_{t-1,im}$ and define $\alpha'_{1mn} = \phi_{mn}(y_1)u_m v_n$. From Eq.\eqref{bb}, we then have $\alpha_{tmn} \propto \alpha'_{tmn}, \forall t$. Define $c_t = \sum_{m=1}^M \sum_{n=1}^N \alpha'_{tmn}$. It can be shown that $\alpha_{tmn} = \alpha'_{tmn}/c_t$. Denote the matrix $\textbf{A}_t = (\alpha_{tmn})_{M \times N}$, we can show that

\begin{equation}\label{matrixfiltering}
\textbf{A}_t = \frac{1}{c_t} \Phi_t \circ (\textbf{W}\textbf{A}_{t-1}), \quad t>2
\end{equation}

\noindent where $\circ$ denotes the \emph{entrywise} (or Hadamard) product of the matrix. $\Phi_t$ and $\textbf{W}$ denote the emission matrix and transtion matrix, respectively, as described in Section 2. For the initial case, we have

\begin{equation}\label{matrixfiltering2}
\textbf{A}_1 = \frac{1}{c_1} \Phi_1 \circ (\textbf{u}\textbf{v}^T)
\end{equation}

\noindent where $\textbf{u}$ and $\textbf{v}$ are as defined in Section 2. To get a matrix formula for a smoothing density, we first define

\begin{equation}
\beta_{tmn} = \frac{p(y_{t+1}^{T}|z_{t}=a_{m},PE^{*}=b_{n})}{p(y_{t+1}^T|y_1^t)}.
\end{equation}

\noindent From Eq.\eqref{cc}, we then have the smoothing density for $t < T$
\begin{equation}
p(z_{t}=a_{m},PE^{*}=b_{n}|y_{1}^{T}) = \alpha_{tmn}\beta_{tmn}.
\end{equation}

\noindent Denote the matrix $\textbf{B}_t = (\beta_{tmn})_{M \times N}, t<T$, we can show that

\begin{equation}\label{matrixsmoothing1}
\textbf{B}_{T-1} = \frac{1}{c_T} \textbf{W}^T\Phi_T,
\end{equation}

\noindent and
\begin{equation}\label{matrixsmoothing2}
\textbf{B}_{t} = \frac{1}{c_{t+1}} \textbf{W}^T(\Phi_{t+1} \circ \textbf{B}_{t+1}), \quad t \in \{1,...,T-2\}.
\end{equation}

\subsection{Inference with unknown parameters}\label{subsec3.2}

In general situations, $\theta$ is unknown, so only the observed data $y_{1}^{T}$ is available. In this case, $\theta$ must be estimated first. \emph{Expectation Maximization} (EM) is a general method capable of estimating the parameters $\theta$ in\emph{ Maximum Likelihood} and \emph{Maximum a Posteriori} (MAP) problem settings for probabilistic models with latent variables \cite{dempster1977maximum}. Here, we formulate our parameter estimation in the MAP setting so that expert's prior knowledge can be employed into the model. Formally, we would like to solve the following problem of maximizing the posterior pdf of $\theta$.
\begin{equation}\label{ee}
  \theta_{MAP}=\operatorname*{arg\,max}_{\theta\in\Theta}(\theta|y_{1}^{T}).
\end{equation}

EM find a solution of Eq. (\ref{ee}) by iteratively solving the following two steps with the arbitrary set of initial parameters $\theta^{(1)}$ and a prior $p(\theta)$. Iterating from $j = 1,2,...$, do

\textbf{E-Step}: Calculating smoothing probabilities $p(z_{t}=a_{m},PE^{*}=b_{n}|y_{1}^{T},\theta^{(j)})$, $\forall t,m,n$

\textbf{M-step}: Solving the constraint maximization problem,
\begin{equation}\label{ff}
  \theta^{(j+1)}=\operatorname*{arg\,max}_{\theta\in\Theta} [Q(\theta;\theta^{(j)})+\ln p(\theta)]
\end{equation}
where
\begin{equation}\label{gg}
 Q(\theta;\theta^{(j)})=E_{z_{1}^{T},PE^{*}|y_{1}^{T},\theta^{(j)}} [\ln p(y_{1}^{T},z_{1}^{T},PE^{*}|\theta)]
\end{equation}

EM repeats the two steps until $\theta^{(j)}$ converges. Note that EM guarantees to find a local maxima of Eq. (\ref{ee}) \cite{bk:Bishop2006}. The argument in the expectation of Eq. (\ref{gg}) is simply the log-likelihood of the model:
\begin{equation}
  \ln p(y_{1}^{T},z_{1}^{T},PE^{*}|\theta)=\sum_{t=1}^{T} \ln p(y_{t}|z_{t},PE^{*},\theta)+\sum_{t'=2}^{T} \ln p(z_{t'}|z_{t'-1},\theta) +\ln p(z_{1}|\theta) +\ln p(PE^{*}|\theta).
\end{equation}

According to DBN, they are simply the logarithms of the emission pdf, transition pdf and initial pdf, respectively. By equation manipulations, the expectation Eq. (\ref{gg}) can be calculated by employing the smoothing probabilities already done in the E-step. As a result, we get a closed form of Eq. (\ref{gg}). Combining with the $\ln p(\theta)$ term described below, the constraint maximization Eq. (\ref{gg}) is well defined and readily to be solved by using the method of \emph{Lagrange multipliers} \cite{boyd2007tutorial}. All derivations details, which have the same mathematical structure for the simpler case of HMM \cite{bk:Bishop2006}, are quite long and can be found in the full version of this paper \cite{bk:Haizhen2017}.

Experts can put their knowledge into the parameter estimation procedure via $p(\theta)$ in Eq. (\ref{ff}).
Here, we assume that all parameters are independent, $p(\theta)=p(\sigma)p(\textbf{u})p(\textbf{v})p(\textbf{W})$. In our experience, investment experts usually have \emph{two} types of knowledge which are useful to estimate $\theta$. The first type of knowledge is about $PE^{*}$. Often, experts may be able to estimate \emph{the range of appropriate $PE^{*}$ level} by analyzing a firm's business strategy together with competitions in its industry. The second type of knowledge is about the \emph{degree of persistence of the medium-term noisy effect} which makes a stock price deviates from its fundamental for a considerable amount of time as explained in Section \ref{sec2}. For some firms, e.g. a firm with non-existent investor relation department, when there exist some unconfirmed rumors, its price can deviate from its fundamental for a long period. In contrast, some firms with both strong public and investor relation departments can clear up unconfirmed rumors rather quickly, so this rumor effect will not stay long. The two types of expert information can be encoded on $p(\textbf{v})$ and $p(\textbf{W})$, respectively. The prior $p(\textbf{v})$ for the vector $\textbf{v}=(v_{n})_{N \times 1}$ can be represented via the \emph{Dirichlet distribution}:
\begin{equation}\label{aaa}
  p(\textbf{v})=\frac{\tau({k_{1}+k_{2}+...+k_{N}})}{\tau({k_{1}})\tau({k_{2}})...\tau({k_{N}})}\prod_{n=1}^{N}v_{n}^{k_{n}-1}
\end{equation}
Intuitively, $k_{n}, n\in \{1,...,N\}$ is a degree of belief for each possible $PE^{*}$ value $b_{n}$. Experts can employ their believes that some value of $PE^{*}$, e.g. $b_{i}$ is relatively more probable than other values by giving $k_{i}$ relatively higher value than other $k_{n}, n\neq i$. See \cite{bk:GelmanEtal2003} for more details on the Dirichlet prior. The prior on transition matrix $p(\textbf{W})$ encoding the average persistence degree of the medium-term noisy effect can also be described by a product of Dirichlet priors: $p(W)=\prod_{m=1}^{M} p(\textbf{w}_{m})$, where as defined in Section \ref{sec2}, $\textbf{w}_m=(w_{im})_{i=1,...,M}$ denotes a $M\times1 $ vector of a probability $p(z_{t+1}=a_{i}|z_{t}=a_{m}), i=1,...,M $, and
\begin{equation}\label{bbb}
  p(\textbf{w}_{m})=\frac{\tau(\sum_{i=1}^{M}k_{im})}{\prod_{i=1}^{M}\tau(k_{im})}\prod_{i=1}^{M} w_{im}^{k_{im}-1}
\end{equation}
The persistence degree of the medium-term effect can be set by relatively increasing the values of $k_{mm}$ compared to other values $k_{im}, i \ne m$. The relatively higher of $k_{mm}$, the more persistence of the medium-term effect. Since medium-term effect appears at random, other values can symmetrically be set: $k_{im}=(1-k_{mm})/(M-1)$, for $i\neq m$.

\section{Experiments}\label{sec4}

In this section, we illustrate benefits of our methodology in real-world applications. To do this, we will conduct comprehensive trading simulations to show consistent superior performances of our method over standard benchmark. In literatures of Finance, according to \emph{Efficient Market Hypothesis} \cite{bk:Fama1976,Fama1991,bk:Mark2011} which is widely accepted by mainstream researchers, the gold standard benchmark is, surprisingly, the simple \emph{``buy and hold''} method which empirically proves to be efficient in the long run. Astonishingly, many evidences clearly indicate that most mutual fund managers who apply complex active portfolio management techniques cannot beat the simple ``buy and hold'' of the market portfolio \cite{bk:Malkiel2016,Carhart1997}. In this paper, we will test our method against this gold standard ``buy and hold'' method both on individual stock level and on portfolio level.

\subsection{Experiment setting}\label{subsec4.1}

In this paper, we will make trading simulations in the markets of two different countries where we can access historical data: NYSE (New York Stock Exchange) and NASDAQ in US and SET (Stock Exchange of Thailand) in Thailand. While NYSE and NASDAQ represent matured stock markets, SET represents an emerging market, so that we are able to test our methodology to firms on both market phases. For each country, we collect 10 firms from various industries to ensure that our methodology is not just limited to one specific industry. Each selected firm is well established and has at least 5 year historical trading data. The names of selected companies with their respective sectors for Thai stocks and US stocks are shown in Table \ref{tab:aa} and Table \ref{tab:bb}, respectively.

We collect daily 5-year historical closing-price data for each firm (Jan 1, 2012 to Sep 30, 2016) consisting of 1160 closing prices for stocks in SET and 1195 closing prices for stocks in NYSE and NASDAQ, respectively. The difference in the number of data is due to different working days in the two countries.
All data are adjusted for \emph{stock splitting} and \emph{stock dividending} if occuring during this 5-year period. To avoid duplicated writing, here, we shall explain only experiment settings for stocks in SET with historical price $P_{1},...,P_{1160}$.The experiment settings for stocks in NYSE and NASDAQ are done similarly.

\begin{table}[!ht]
\vspace{0.1 cm}
\begin{center}

\caption{Stock symbols of selected firms from Thailand, together with their industries. BigCap, MidCap and SmallCap are defined according to market sizes (in THB) which are greater than 100 billions, 10 billions and 1 billions, respectively.}\label{tab:aa}\footnotesize
\end{center}
\begin{center}
\vspace{0 cm}
\begin{tabular}{ l l l }
\hline
\linespread{1.6}

Symbol & Industry  &\ \ \  Size  \\
\hline
CPALL  & Retailing-Food and Staples  & BigCap   \\

CPN  & Developer-Department Stores  & BigCap  \\

EASTW  & Utilities-Water Resources  & MidCap   \\

GLOW  & Utilities-Power Plant  & BigCap   \\

HMPRO  & Retailing-Household Products  & BigCap   \\

QH  & Developer-Housing  & MidCap   \\

ROBINS  & Retailing-General  & MidCap   \\

SCB  & Banking  & BigCap   \\

SNC & Electrical Equipments  & SamllCap   \\

TTW & Utilities-Tap Water  & MidCap   \\

\hline

\end{tabular}
\vspace{-.5 cm}
\end{center}
\end{table}

\begin{table}[!ht]
\vspace{0.1 cm}
\begin{center}
\caption{Stock symbols of selected firms from US, together with their industries. BigCap, MidCap and SmallCap are defined according to market sizes (in USD, Data retrieved on Dec 7, 2016)}\label{tab:bb}\small
\end{center}
\begin{center}
\vspace{0 cm}
\begin{tabular}{ l l l }
\hline
Symbol & Industry  &\ \ \  Size  \\
\hline
WMT  & Services-Discount and Variety stores  & $218.48B$   \\

HD  & Services-Home improvement stores  & $157.86B$   \\

KO  & Consumer Good-Beverages soft drinks  & $173.51B$   \\

G  & Services-Business services  & $4.925B$   \\

AAPL  & Consumer Goods-Electronic equipment  & $584.37B$   \\

NKE  & Consumer Goods-Textile-Apparel Footwear and Accessories  & $84.35B$   \\

BK  & Financial-Asset Management  & $51.89B$   \\

CF  & Basic Materials-Agricultural Chemicals  & $6.695B$   \\

CSCO & Technology- Networking and Communication Devices  & $147.88B$   \\

DIS & Services-Entertainment Diversified  & $157.41B$   \\

\hline

\end{tabular}
\vspace{-.5 cm}
\end{center}
\end{table}

For each firm, the corresponding yearly earnings data in those years are also collected. $E_{1},...,E_{1160}$ are defined as the summation of the most recent 4 quarterly earnings on each data $t$. The first 3-year historical data (Jan 1,2012 to Dec 31,2014) $P_{1},...,P_{735}$ and $E_{1},...,E_{735}$ will be used as a training data for our Bayesian methodology to learn the appropriate parameters $\theta=\{\textbf{W,u,v},\sigma^{2}\}$ using the EM algorithm as well as estimate the most probable values of $PE^{*}$ and $\{z_{1},...,z_{735}\}$ by the method of smoothing as explained in Section \ref{sec3}. The constants $\{a_{1},...,a_{M}\}$ and $\{b_{1},...,b_{N}\}$ are set by experts. Since the constant $M$ determines the size of the transition matrix $\textbf{W}=(w_{im})_{M\times M}$, we make a constraint $M<10$ so that the model is not over-parameterized and that 3-year historical data is enough to learn $\textbf{W}$. For all prior distributions, we employ non-informative priors with the exception of $p(\textbf{W})$ where our ``security experts'' emphasize the prior knowledge of $z_{t}$ \emph{persistency} as described in Subsection \ref{subsec3.2}.

Each trading simulation is conducted for each individual stock with the remaining 2-year historical data $P_{736},...,P_{1160}$ to measure the performance of both our method and the benchmark. The performance measurement metric is, as used by practitioners, \emph{a profit generated by each method}. The profit calculation is straightforward: for each trading simulation, each method is equally given an initial amount of cash $I$ to make a trade (which taken into account a commission fee), and the profit are simply all the asset values at the end of a simulation minus $I$. For simplicity, we assume that each stock can be bought with all the money we have, e.g. supposing we have  $100\$$ and  a  stock's price is  $12\$ $, then we are able to buy $100/12=8.33$ stocks.

Using the benchmark ``buy and hold" strategy we simply need to buy the stock with all the cash at the beginning and then do nothing until the end. Initially, this method will get $C.I/P_{736}$  shares where $C\approx0.9987$ represents the value of assets after taking SET's commission fee into account. At the end of the simulation this asset will have a value of $P_{1160}.C.I/P_{736}$, so the profit can be calculated easily. For a trading strategy employed by our method, there are two possible versions inspired by our model's main idea (see Figure \ref{fig:Fig1}) and\emph{ Strategy A} (buy low, sell high) described in Section \ref{sec1}. The first version called \emph{long-term strategy} is simply to ``buy low, sell high" with respect to the static value of $PE^{*}$, and the second version called \emph{medium-term strategy} is to ``buy low, sell high" with respect to the dynamic values of $PE^{*}(1+z_{t})$ where each $z_{t}$ is dynamically estimated by the method of filtering described in Subsection \ref{subsec3.2}. Both versions can be formally described as follows.

Let $I_{t}$ and $N_{t}$ be available cash and total shares at date $t$, respectively. Initially, $I_{736}=I$ and $N_{736}=0$. Now, both trading versions can be defined simply by the following procedure: for each date $t$, exactly one of the following cases holds:

(i) $P_{t}/E_{t}\leq A_{t}(1-Tr)$ and $I_{t}>0$ (buy-low case) where $Tr\in (0,1)$ is a threshold, $A_{t}=PE^{*}$ for the long-term strategy and $A_{t}=PE^{*}(1+z_{t})$ for the medium-term strategy. In this case, buy the stock with all cash, so that $N_{t+1}=C.I_{t}/P_{T}$ and $I_{t+1}=0$.

(ii) $P_{t}/E_{t}\geq A_{t}(1-Tr)$ and $I_{t}=0$ (sell-high case). In this case, sell all the holding stock to get cash $I_{t+1}=P_{t}.N_{t}.C$ and $N_{t+1}=0$.

(iii) If case (i) and case (ii) are not satisfied, do nothing. So, $I_{t+1}=I_{t}$ and $N_{t+1}=N_{t}$.
At the end of a trading simulation $t=1160$, the total profit is simply $I_{1160}+P_{1160}.N_{1160}-I_{736}$, so that we can compare with the "buy and hold" profit.

\begin{figure}[!htb]
  \centering

  \includegraphics[width=0.6\linewidth]{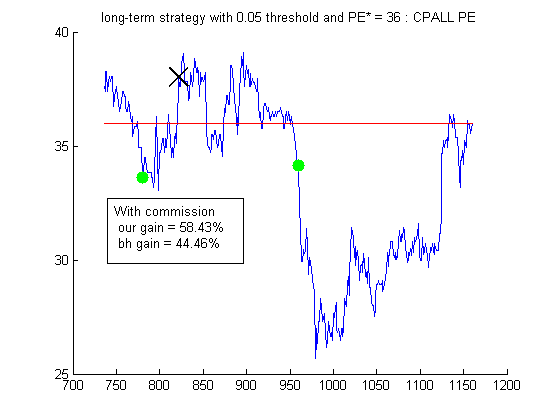}
  \includegraphics[width=0.6\linewidth]{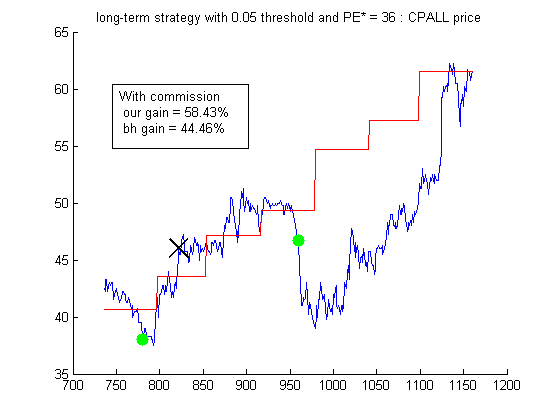}

  \caption{CPALL example of \emph{long-term strategy} trading with threshold $Tr = 5\%$ where our model's profit is 58.43\% while ``buy \& hold'' profit is 44.46\%. ``Green circle'' denotes ``buy'' and ``Black cross'' denotes ``sell''. Top and bottom figures show the same trading in different perspectives The top figure shows trading with respect to the ``PE'' perspective where Red line denotes $PE^{*}$. Here, it is easy to see our strategy in action: when the observed PE is lower or higher than the threshold level, buy or sell is triggered, respectively. The bottom figure shows trading in the ``price'' perspective when Red line denotes $P^{*}=E_{t}PE^{*}$. Since the earnings continue to increase, $P^{*}$ also increase accordingly.}

  \label{fig:Longterm}

\end{figure}

\begin{figure}[!htb]
  \centering

  \includegraphics[width=0.6\linewidth]{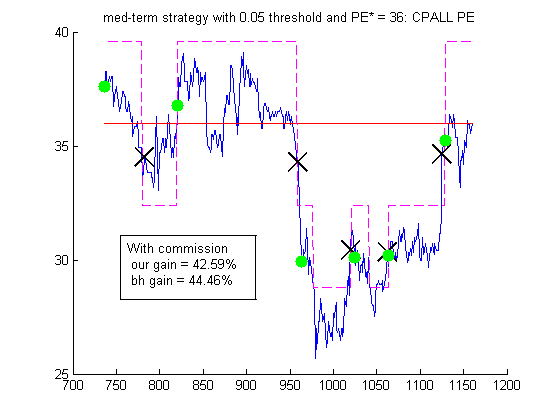}
  \includegraphics[width=0.6\linewidth]{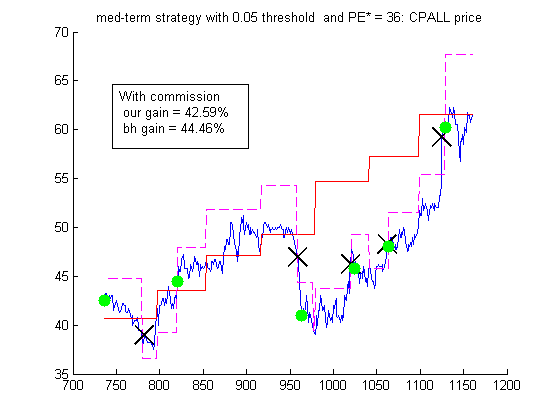}

  \caption{Example of\emph{ medium-term strategy} trading of CPALL with threshold  $Tr = 5\%$. In addition to those explained in Figure \ref{fig:Longterm}, in the top figure, the purple dashed line denotes $PE^{*}(1+z_{t})$, and becomes the base line of this trading strategy. Note that our Bayesian method estimates the purple line by the method of ``filtering'' which tracks the observed PE movements with some delay. The bottom figure is the ``price'' perspective where the purple line shows $P^{*}(1+z_{t})$. In this example, ``buy \& hold'' method beats ours by small margin because of the commission fees caused by our frequent trading.}

  \label{fig:Medterm}

\end{figure}

We give some illustrations of our trading in actions which are shown in Figure \ref{fig:Longterm} and Figure \ref{fig:Medterm}. Figure \ref{fig:Longterm} is an example of long-term trading of CPALL with threshold 0.05 and Figure \ref{fig:Medterm} is an example of medium-term trading of CPALL with threshold 0.05.

\subsection{Experimental results and discussions} \label{subsec4.2}

\subsubsection{Individual-firm level experiments}

To ensure that our experimental results are not biased because of a threshold choice, we test 4 different thresholds for each trading strategy. Note that the thresholds in the medium-term trading are relatively smaller than those in the long-term. This is due to the nature of medium-term strategy where $PE_t$ has a smaller deviation from its base line $PE^*(1+z_t)$ compared to the long-term strategy's base line, not containing the effect of $z_t$. The experimental results with respect to Thai stocks and US stocks are shown in Table \ref{tab:cc} and Table \ref{tab:dd}, respectively.
\begin{landscape}
\begin{table}[htbp]
\centering
\caption{Experimental results in profit percentage of our Bayesian trading strategies compared to the benchmark ``buy and hold'' method on Stock Exchange of Thailand (SET). Bold face numbers indicate a case which our method is superior. On the last row, ``W/D/L'' summarizes Win/Draw/Lose of our method compared to the benchmark. }\label{tab:cc}
\begin{tabular}{l|rrrr|rrrr|r}
\hline
\multirow{ 2}{*}{Symbol} & \multicolumn{4}{c|}{Long-term Thresholds} &\multicolumn{4}{c|}{Medium-term Thresholds}& \multirow{ 2}{*}{Buy \& Hold}\\ \cline{2-9}
& 5\%  &10\% &15\% &20\%  & 3\%&5\%&7\% &10\% & \\
\hline

CPALL  &\textbf{58.40\%} &38.74\% &\textbf{49.74\%} &\textbf{56.42\%} &26.92\% &42.59\% &32.08\% &36.34\% &44.46\%  \\

CPN  &30.67\% &30.67\% &30.67\% &30.67\% &-1.06\% &-6.39\% &-4.32\% &14.04\% &30.67\%   \\

EASTW &9.16\% &6.18\% &4.29\% &4.29\% &\textbf{10.49\%} &3.93\% &\textbf{15.07\%} &\textbf{20.64\%} &9.16\%  \\

GLOW  &-11.78\% &-11.78\% &-11.78\% &-11.78\% &\textbf{-3.24\%} &-14.16\% &\textbf{-10.66\%} &\textbf{-3.13\%} &-11.78\%  \\

HMPRO  &31.88\% &\textbf{47.93\%} &\textbf{58.00\%} &0\% &\textbf{51.69\%} &\textbf{69.87\%} &\textbf{86.49\%} &8.87\% & 32.28\%  \\

QH  &-24.08\% &-24.08\% &-24.08\% &\textbf{-9.70\%} & \textbf{5.73\%} & \textbf{0.49\%} &\textbf{4.08\%} &\textbf{2.31\%} &-24.08\%  \\

ROBINS  & 36.28\% & 36.28\% &36.28\% &36.28\% &32.70\% &\textbf{36.77\%} &\textbf{45.64\%} &\textbf{36.97\%} &36.28\%  \\

SCB  &-17.46\% & -17.46\% &\textbf{-14.10\%} &\textbf{-9.08\%} &\textbf{-11.67\%} &\textbf{-14.09\%} &\textbf{-14.74\%} &\textbf{4.79\%} &-17.46\% \\

SNC & -4.45\% &-4.45\% &-7.00\% &\textbf{-0.91\%} &\textbf{18.46\%} &\textbf{8.20\%} &\textbf{8.02\%} &\textbf{8.20\%} &-4.45\%   \\

TTW & \textbf{0.00\%} &\textbf{0.00\%} &\textbf{0.00\%}  &\textbf{0.00\%} &\textbf{-1.31\%} &\textbf{-0.50\%} &\textbf{3.67\%} &\textbf{0\%} &-3.73\%  \\
\hline
\textbf{Average} & \textbf{10.86\%} &\textbf{10.20\%} &\textbf{12.19\%}  &\textbf{9.62\%} &\textbf{12.92\%} &\textbf{12.67\%} &\textbf{16.55\%} &\textbf{12.90\%} &9.14\%  \\
\hline

\multirow{ 2}{*}{\textbf{W/D/L}}
& 2/7/1 & 2/6/2 & 4/4/2 & 5/3/2

& 7/0/3 & 6/0/4 & 8/0/2 & 7/0/3 \\ \cline{2-9}

& \multicolumn{4}{c|}{13/20/7} & \multicolumn{4}{c|}{28/0/12} & \\
\hline

\hline

\end{tabular}
\end{table}
\end{landscape}
%
%
%
%

\begin{landscape}
\begin{table}[htbp]
	\centering
	\caption{Experimental results in profit percentage of our Bayesian trading strategies compared to the benchmark ``buy and hold'' method on US stock markets: NYSE and NASDAQ.}\label{tab:dd}
\begin{tabular}{l|rrrr|rrrr|r}
\hline
\multirow{ 2}{*}{Symbol} & \multicolumn{4}{c|}{Long-term Thresholds} &\multicolumn{4}{c|}{Medium-term Thresholds}& \multirow{ 2}{*}{Buy \& Hold}\\ \cline{2-9}
& 5\%  &10\% &15\% &20\%  & 3\%&5\%&7\% &10\% & \\
\hline

		WMT  &\textbf{6.07\%} &\textbf{13.00\%}&\textbf{23.04\%} &\textbf{0.00\%} &\textbf{20.47\%} &\textbf{5.97\%} &\textbf{12.02\%} &\textbf{0.00\%}  &-11.98\%  \\
		
		HD  &10.56\% &28.71\% &28.71\% &28.71\% &\textbf{48.35\%} &\textbf{40.77\%} &27.55\% &27.48\% &28.71\%   \\
		
		KO &6.06\% &6.06\% &6.06\% &6.06\% &\textbf{20.61\%} &\textbf{17.24\%}  &0.00\% &0.00\% &6.06\%  \\
		
		G & 5.33\% &26.50\% &26.50\%  &26.50\% &\textbf{37.40\%} &\textbf{29.74\%}  &21.22\% &1.14\% &26.50\%  \\
		
		AAPL  &5.19\% &5.19\% &5.19\% &5.19\% &\textbf{18.69\%} &\textbf{22.83\%} &\textbf{41.27\%} &\textbf{26.84\%} & 6.77\%  \\
		
		NKE  &12.72\% &12.72\% &12.72\% &12.72\% &\textbf{22.53\%} &\textbf{25.00\%} &\textbf{24.50\%} &\textbf{32.96\%} & 12.72\%  \\
		
		BK  &\textbf{26.36\%} &\textbf{23.72\%} &1.42\% &1.42\% &\textbf{5.52\%} &\textbf{6.44\%} &-3.68\% &\textbf{4.84\%} &1.42\%  \\
		
		CF  &\textbf{27.16\%} &\textbf{15.87\%}   &-53.74\% &-53.74\% &\textbf{-17.60\%} &\textbf{-16.84\%} &\textbf{-36.21\%} &\textbf{-41.05\%} &-53.74\%  \\
		
		CSCO  &12.00\% &\textbf{23.00\%} &20.48\% &20.48\% &8.42\% &10.98\% &14.98\% &20.00\% &20.48\% \\
		
		DIS & 0.82\% & 0.82\% & 0.82\% & 0.82\% &\textbf{14.37\%} &\textbf{12.17\%} &-1.45\% &-8.47\% & 0.82\%   \\

		\hline
\textbf{Average} & \textbf{11.23\%} &\textbf{15.56\%} &\textbf{7.12\%}  &\textbf{4.82\%} &\textbf{17.88\%} &\textbf{15.43\%} &\textbf{10.02\%} &\textbf{6.37\%} &3.78\%  \\
		\hline
		\multirow{ 2}{*}{\textbf{W/D/L}}
		& 3/3/4 & 4/5/1 & 1/8/1 & 1/8/1
		
		& 9/0/1 & 9/0/1 & 4/0/6 & 5/0/5 \\ \cline{2-9}
		
		& \multicolumn{4}{c|}{9/24/7} & \multicolumn{4}{c|}{27/0/13} & \\
		\hline

		\hline
	\end{tabular}
\end{table}
\end{landscape}

From the tables, we can see that in the total of 80 trading simulations on SET firms, our method results in greater performance 41 times, while ``buy and hold" results in better performance 19 times (the remaining 20 times are draws). Similarly, in the total of 80 trading simulations on NYSE and NASDAQ firms, our method results in greater performance 36 times, while ``buy and hold" results in better comparison 20 times (the remaining 24 times are draws). Summing up results of markets in the two countries, our method outperforms the benchmark 77 times, yet underperforms only 39 times. These are promising results where we shall analyze \emph{statistically significancy} of the results more formally in the next subsection. Here, we shall firstly interpret and discuss about experimental results in Tables \ref{tab:cc} and \ref{tab:dd} in details.

From Tables \ref{tab:cc} and \ref{tab:dd}, it can be seen that there are 44 draws, which occur only in the cases of the long-term trading strategy. All 44 draws happen because of the exact same reason: \emph{our model predicts of undervaluation at the beginning of the testing period}. i.e. the \emph{first} \emph{observed PE} falls deeply below the base line $PE^*$ (exceeding the specified threshold). \emph{After that, the observed PE is never once able to overly exceed $PE^*$ with respect to the given threshold}, i.e. $PE^*(1+Tr)$ is quite high in the test set so that no selling is possible. Therefore, in this case, our trading behaves exactly just like the benchmark ``buy and hold". Note that there is no draw in the results of the medium-term trading strategy. The reason is because the estimated medium-term noisy effect $z_t$ makes the base line $PE^*(1+z_t)$ move near to the \emph{observed PE} which results in more frequent trading.

\begin{figure}[!htb]
	\centering
	
	\includegraphics[width=0.8\linewidth]{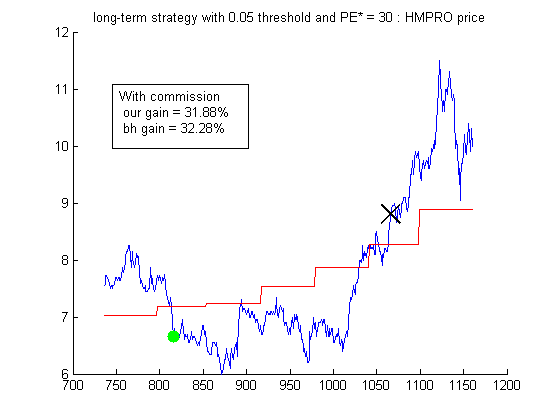}
	
	\caption{Example of a \emph{growth stock} \cite{bk:LynchRothchild2000} where its earnings are consistently increasing. For a growth stock, if the threshold size is non-optimal, we may buy and sell too early (and do not make trading frequently enough) which results in less profit.}
	
	\label{fig:HmproLT5}
	
\end{figure}

\begin{figure}[!htb]
	\centering
	
	\includegraphics[width=0.8\linewidth]{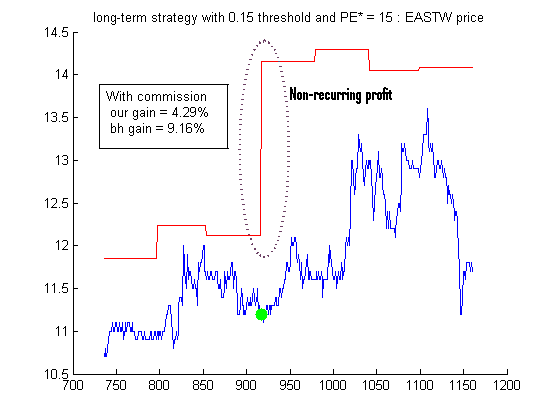}
	
	\caption{Example of \emph{non-recurring earnings} or \emph{non-recurring profit}. Since our fundamental price calculation is naive and does not take into account the fact that these earnings are temporary, it estimates a too-high fundamental price due to this extra earnings.}
	
	\label{fig:Eastw}
	
\end{figure}

Disregarding the draws, our long-term trading strategy still beats the benchmark with 22 wins versus 14 loses. This is mainly due to the volatility of the observed PE in most stocks so that our strategy of buying in an undervalued price and selling in an overvalued price with respect to $PE^*$ is possible. However, it is not the case that trading induced from our model constantly outperforms the benchmark. For the case of the so-called \emph{growth stocks} \cite{bk:LynchRothchild2000}, i.e. stocks with consistently increasing earnings and price, it is not so easy for our model to beat the benchmark. If the threshold is set too low, our method will lead to buying and selling early, and thus results in less profit. See Figure \ref{fig:HmproLT5} for example. If the threshold is set too high, our method may lead to buying when the price is already high or lead to doing nothing at all because it is never undervalued with respect to the specified threshold. Another special case where our method fails to beat the benchmark is when there are the so-called \emph{non-recurring} earnings, i.e. extra incomes which occur only once and should not be taken into account the calculation of fundamental value. In this case, the market knows that these extra earnings are temporary and do not give it credit, i.e. the price does not go up according to this profit. Our method has not taken this information into account and thus is fooled to believe that a stock is undervalued. See Figure \ref{fig:Eastw}.

On the other hand, The results of our method equipped with medium-term trading strategy show impressive superiority, 55 wins versus 25 loses to the benchmark. The key factor of success  is its tracking ability of the medium-term noisy effect $z_t$ by our filtering algorithm presented in Section \ref{sec3}. When the new base line $PE^*(1+z_t)$ is predicted accurately, undervalued and overvalued prices are also accurately detected and so the probability of our profitable trading is increasing. Nevertheless, with more frequent trading, commission fees increase substantially and sometimes can significantly reduce our performance as shown in Figure \ref{fig:Medterm}.

\subsubsection{portfolio level experiments}

In this subsection, to more realistically simulate a real-world individual investor, we construct a \emph{portfolio of stocks} and test its performance against the benchmark. Here, we use a rule-of-thumb commonly employed in practice saying that a good portfolio should consists of around 15 stocks \cite{Investopedia,EffFront} (which  contradicts to the mainstream theory \cite{domian2007diverse}). For individual value investors who believe they can beat the market by analyzing each firm carefully, usually, they do not feel comfortable to hold too many stocks (like $100$ stocks recommended by academic financial literatures) because investors need time to update and analyze the information of all their holding stocks.

To test the performance of a 15-stock portfolio of our method against the benchmark, we employ the method of \emph{boostrap resampling} \cite{horowitz1997bootstrap}. For each boostrap sample, a set of 15 stocks are selected randomly from Tables \ref{tab:cc} and \ref{tab:dd} to form an \emph{equally-weighted} portfolio. We are interested in the difference in performance between our method and the benchmark on each boostrap sample. After all bootsrap samples are drawn, we can also estimate the \emph{average} difference in performance between the two methods. More precisely, let $X$ be a random variable representing difference in \emph{\% profit} between our model and the benchmark (our \emph{\% profit} minus the benchmark's \emph{\% profit}). By repeating the boostrap resampling \emph{10,000 times}, we are able to construct an \emph{empirical distribution} of $X$. This empirical distribution allows us to calculate $E[X]$, the average \emph{\% profit} difference between the two methods, and $Pr(X \ge 0)$, the probability that our method has superior or equal performance to the benchmark. In addition to a portfolio consisting of stocks from the two markets, we also test portfolio performance from SET or US alone. Since the number of stocks considered in this work shown in Tables \ref{tab:cc} and \ref{tab:dd} is 10, a 7-stock portfolio is constructed for these cases instead of a 15-stock portfolio. The results are shown in Table \ref{tab:port}.

\begin{table}[t]
	\centering
	\caption{Experimental results in portfolio level testing. $X$ denotes a random variable representing difference in \emph{\% profit} between our model and the benchmark. The distribution of $X$ is estimated using the method of \emph{Boostrap Resampling}. Bold face denotes a case where there is more than 80\% confidence that our method is more superior or equal to the benchmark.}\label{tab:port}\footnotesize
	\begin{tabular}{l|rrrr|rrrr}
		\hline
		\multirow{ 2}{*}{Portfolio} & \multicolumn{4}{c}{Long-term Thresholds} &\multicolumn{4}{c}{Medium-term Thresholds}\\ \cline{2-9}
		& 5\%  &10\% &15\% &20\%  & 3\%&5\%&7\% &10\% \\
		\hline
		
		\underline{\textbf{SET}} \\
		$E[X]$  & \textbf{1.70}\% &	1.07\% &	\textbf{3.12\%} &	0.45\% &	3.73\% &	3.52\% &	\textbf{7.46\%} &	3.70\% 		  \\
		
		$Pr(X \ge 0)$  & \textbf{86.76\%} &	66.72\% &	\textbf{84.99\%} &	57.45\% &	71.16\% &	69.53\% &	\textbf{81.44\%} &	74.28\%  \\ \hline
		
		\underline{\textbf{US}} \\
		$E[X]$& 7.30\% &	\textbf{11.88\%} &	3.35\% &	1.03\% &	\textbf{14.04\%} &	\textbf{11.75\%} &	\textbf{6.22\%} &	2.62\%   \\
		
		$Pr(X \ge 0)$ & 72.25\% &	\textbf{97.06\%} &	72.51\% &	72.29\% &	\textbf{99.82\%} &	\textbf{99.78\%} &	\textbf{88.48\%} &	70.35\%   \\ \hline

		\underline{\textbf{SET+US}} \\
		$E[X]$  &	\textbf{4.62\%} &	\textbf{6.42\%} &	\textbf{3.19\%} &	0.77\% &	\textbf{8.89\%} & 	\textbf{7.55\%} &	\textbf{6.87\%} &	\textbf{3.18\%}   \\
		
		$Pr(X \ge 0)$  &	\textbf{80.64\%} &	\textbf{97.47\%} &	\textbf{92.78\%} &	66.04\% &	\textbf{97.94\%} &	\textbf{96.30\%} &	\textbf{92.74\%} &	\textbf{80.58\%}  \\	
		
		\hline
	\end{tabular}
\end{table}

\begin{figure}[!htb]
	\centering
	
	\includegraphics[width=0.8\linewidth]{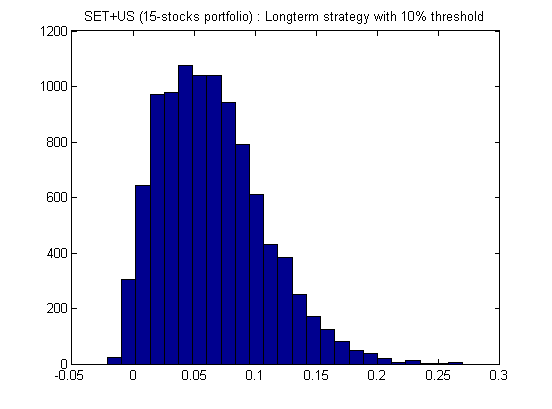}
	
	\caption{A histogram illustrated common empirical distributions of $X$ with right-skewed and long-tail. In this specfic illustration, our method is equipped with the long-term trading strategy with 10\% threshold.}
	
	\label{fig:Histogram}
	
\end{figure}

From Table \ref{tab:port},  our method beats the benchmark on every case \emph{on average} (since $E[X] > 0$ for all cases). However, on a single-country portfolio, about half of the cases have the confidence levels of superiority $Pr(X \ge 0)$ less than 80\%. Most of them have satisfactory confidence levels greater than 70\% though. On the other hand, on a two-country 15-stock portfolio, although $E[X]$ is roughly \emph{an average of the two single-country portfolios}, $Pr(X \ge 0)$ is significantly increasing so that most cases have the confidence levels of superiority greater than $80\%$, half of them is greater than $90\%$. \emph{This statistically confirms the superiority of our method over the benchmark on selected stocks}. This phenomenon of confidence-level increasing is due to the \emph{diversification} effect on portfolio with a higher number of stocks. Finally, we note that the empirical distribution of $X$ is usually \emph{skewed} and \emph{long-tail} as illustrated in Figure \ref{fig:Histogram}. In this non-simple probability distribution case, boostrap empirical-distibution estimation employed in the present paper usually provides more accurate result than traditional analytical asymptotic estimations \cite{horowitz1997bootstrap}.

\section{Conclusion and Future Directions}

In this paper, we propose to apply the advanced \emph{Dynamic Bayesian Network} (DBN) methodology to model \emph{stock price dynamics} with two latent variables, namely, the \emph{fundamental PE} and the \emph{medium-term noisy effect}, respectively. This model is most suitable for one majority category of practitioners, namely, \emph{value investors} in a security market (our model is not suitable for technical investors and mainstream academic investors). We have derived both inference and parameter estimation algorithms. The resulted model can be used as a decision support system to investment experts, or used to construct a trading strategy directly as illustrated in Section \ref{sec4}. Experiments in both individual firm-level and portfolio level show statistically significant superiority of our method.

There are many possible future directions for the present work. The first direction is to more formally reformulate the stock price dynamics model to reflect other important economic and financial variables, e.g. \emph{interest rate, equity risk premium and return of equity}. This modeling process is possible used by the so-called \emph{Gordon Growth Model} \cite{bk:Damodaran2012} which links $PE^*$ to the mentioned variables. \emph{Dynamic Gordon Growth Model} \cite{campbell1988dividend} can be a further future work in this direction. It is also possible to make our DBN model more realistic by allowing time-varying short-term noisy effect, the so-called \emph{dynamic volatility} \cite{wu2013dynamic} or allowing \emph{dynamic volume} \cite{llorente2002dynamic}. Another promising direction in Behavioral finance which can be taken into account in our model is the topic of \emph{heterogeneous agents} \cite{bk:Hommes2013}. To improve our inference procedure, approximate inference such as \emph{Variational Bayes} \cite{bk:Murphy2012}, or stochastic inference such as \emph{Markov chain Monte Carlo} \cite{bk:Bishop2006} are very promising future directions.

\section*{References}

\bibliographystyle{unsrt}
\bibliography{haizhenreference}

\end{document}